\documentstyle[psfig,12pt,blois]{article}

\begin{document}

\heading{\ WHAT CAN WE LEARN FROM TYPE I/II SUPERNOVAE RATES AT HIGH {\sl Z} ?}

\author{R. Sadat $^{1,2}$, A. Blanchard $^{2}$, B. Guiderdoni $^{1}$, J. Silk $^{3}$} {$^{1}$ I.A.P 98bis Bd Arago, 75014 paris Cedex, France}  {$^{2}$ Observatoire de Strasbourg, 11 rue de l'universite 67000 Strasbourg, France\\$^{3}$ Departments of Physics, Astronomy and Center for Particle Astrophysics Astronomy, University of California, Berkeley, USA}

\begin{bloisabstract}
We show how Type Ia/II supernova rates at various redshifts 
can be used to follow the cosmic star formation rate (CSFR) history. Using the recent estimate of the Cosmic Star Formation (CSFR), we compute the predicted cosmic Type Ia/II supernovae rates.
 We show that while SNII rates provide a direct measurement of the instantaneous CSFR, the SNIa rate depends on the past history of the CSFR. This provides us with a new and independent diagnostic of the star formation rate. The comparison with the data are in good agreement with both local and high-$z$ measurement and provides an additional evidence of the validity of the CSFR up to $z = 1$, but suggests higer values at $z > 1$ consistent with the presence of extinction effects. Measurement of SNIa rate up to $z {\sim}$ 1 will be crucial for our understanding of the star formation history at early epochs $z > 1$. We also show that current observations on Type Ia supernovae rate already give us valubale information on the nature of their progenitors.   \end{bloisabstract} 
\section{Introduction}
Recent observations of the most distant galaxies has permitted to probe the history of star formation which is crucial for understanding galaxy evolution. Estimations of the CSFR are however very uncertain because of the poorly known amount of dust absorption and AGNs emissions. Supernova rates at high redshift are very important not only because of their 
ability to allow the direct determination of cosmological parameters 
but also in understanding galaxy evolution and star formation. More specifically, the study of Type Ia supernovae can allow one to probe the past history of star formation, since there is a time delay between the formation of stars and the occurrence of the explosion, whereas their Type II SN counterparts directly probe the instantaneous star formation because of the short-lived massive stars which 
explode. An increasing number of Type Ia SNe, which are the most 
 homogeneous and most luminous SNe, are detected at redshift $z\sim 1$ and  the first measurement of the Type Ia SN rate at $z {\sim} 0.4$ has been recently reported [5].
We use the chemical and photometric population synthesis code of [8] to make predictions on the cosmic evolution of SN rates with 
redshift based on the CSFR derived by [4]. We expect that such studies can provide additional constraints
on the CSFR history and the nature of Type Ia progenitors.
We hereafter adopt $H_{0}$ = 50 kms$^{-1}$Mpc$^{-1}$, ${\Omega}_{0}$ = 1 and ${\Omega}_{\Lambda}$ = 0.
\section{Predicted cosmic evolution of SN Rates as a test to CSFR}
We use our spectrophotometric model of galaxy evolution with a standard IMF which provides a self-consistent description of global properties of local galaxies (see [8] for more details) 
We estimate the rate of Type Ia SNe according to the 
formalism by Ferrini [3]. The free parameters which are the lower limit of the total mass of the systems which can produce a Type Ia SN, and the fraction of the total mass of 
stars which belong to these systems have been adjusted in order to reproduce the main properties of the solar neighbourhood, and specifically the age-metallicity relation in the Milky Way and also the evolution of O and Fe in the solar neighbourhood.
We here restrict ourselves to solar metallicity models. More chemically consistent models are proposed in [8].  
\begin{figure*}[btp]
\centerline{\psfig{file=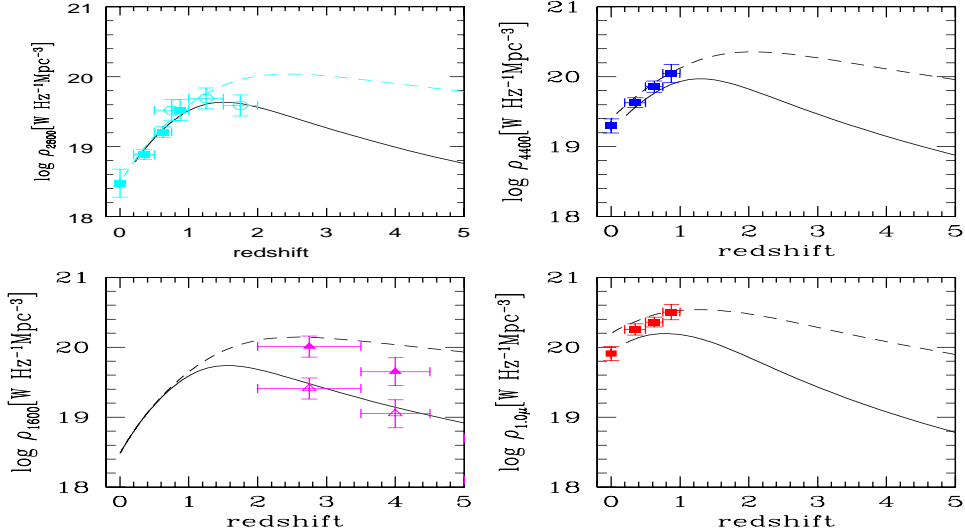,height=3.0in,width=5.3in}}
\caption{\small Evolution with redshift of the luminosity density at 
rest-frame wavelengths. data are taken from [4]. The two curves correspond to the two different parameterizations M1 ({\it solid line}) and M2 ({\it dashed line}) of the CSFR.}
\end{figure*} 
We make use of the CSFR as derived by [4]. We also introduce another CSFR law to account for a possible dust extinction in agreement with [6]. The amplitude is normalized in order to match the luminosity density evolution
in the UV--continuum. Figure 1 shows the predicted luminosity density $\rho_{\nu}$ at various wavelengths. As we can see, the agreement with observations in the other wavelengths is satisfying. The 4400 and 10000 $\AA$ luminosity densities even seem to hint at model M2 in agreement with the IR/Submm background which also suggests the presence of extinction at high $z$. 
\begin{figure*}[btp]
\hbox{\hspace{0cm}\psfig{figure=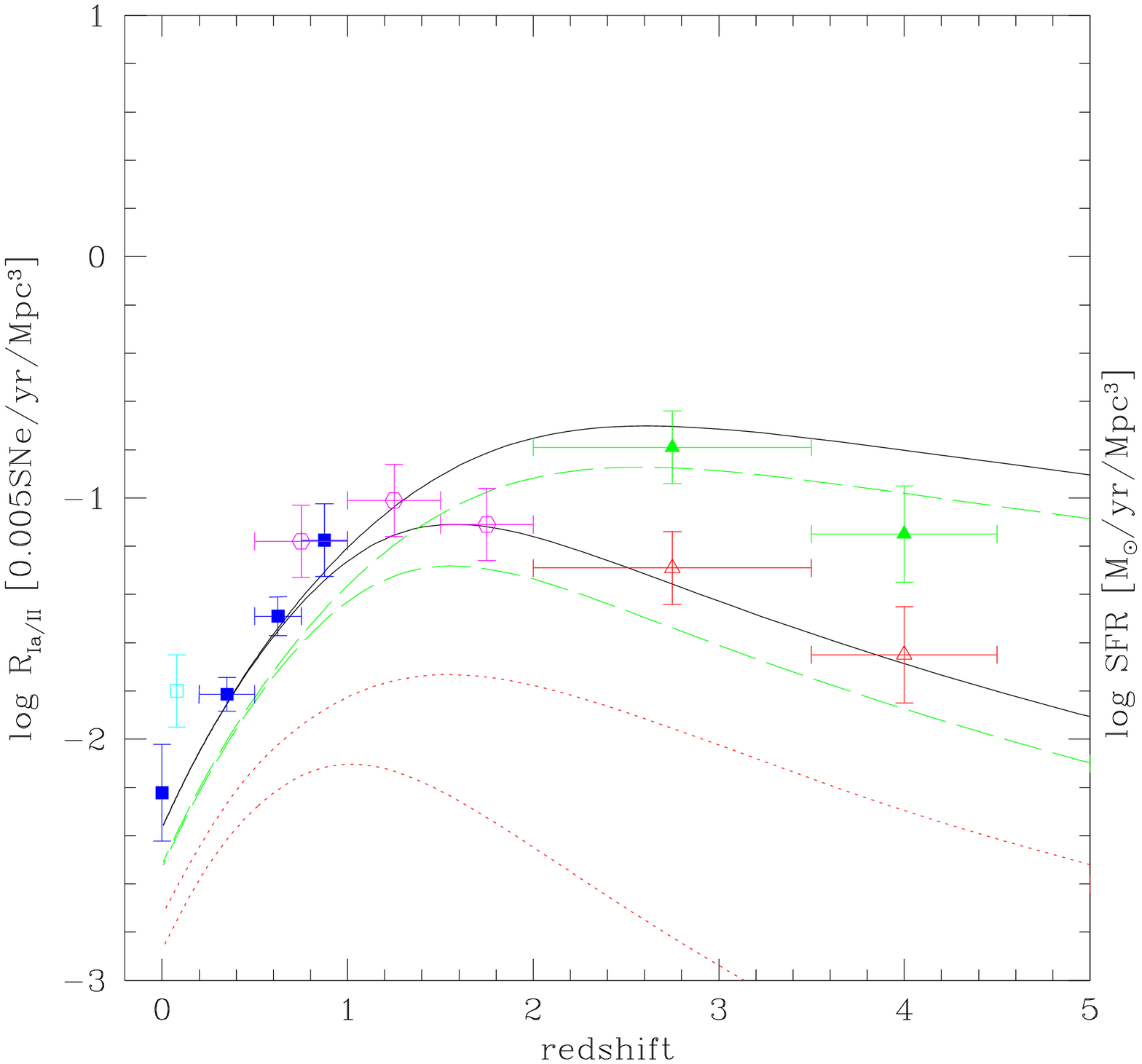,width=8.5cm}\hspace{0cm}
\psfig{figure=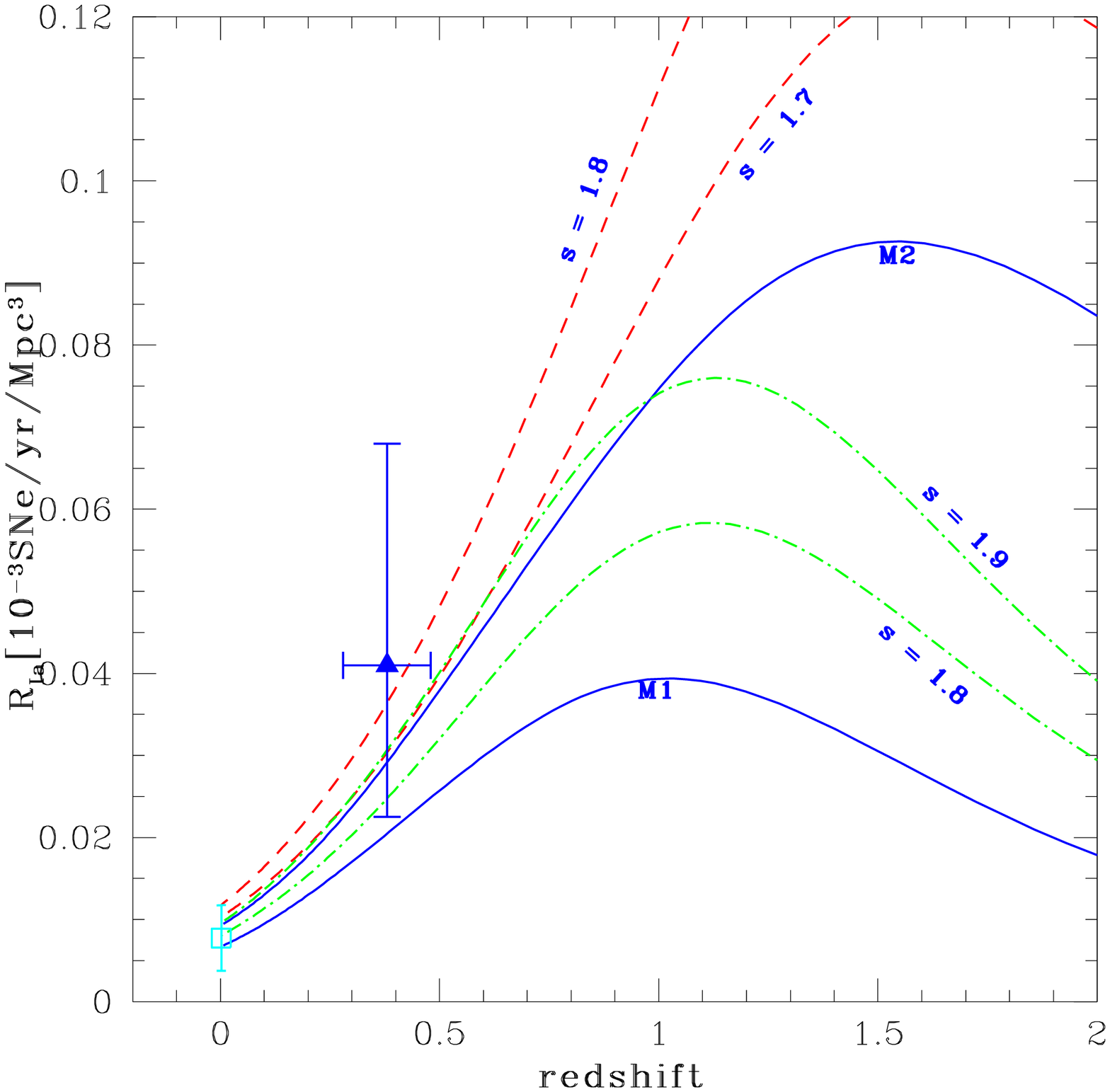,width=8.5cm}}
\caption{{\it left panel}: Evolution of SN rest-frame rates per comoving volume for model M1 ({\it lower curves}) and M2 ({\it upper curves}): $R_{II/Ib,c}$ ({\it solid lines}); ${R_{Ia}}$ ({\it dashed lines}). CSFRs are shown as {\it dotted lines}. {\it right panel}: Evolution of Type Ia SN rest-frame rate per comoving volume for SD models ({\it solid line}) and Ciotti's models ({\it dashed line}). Data are from [1] ({\it empty square}) and [5] ({\it filled triangle}).The horizontal bar is the observational redshift range}
\end{figure*}
\noindent
As already mentionned, the direct measurement of SNe can be used as an independent test for the cosmic star formation in the universe. Figure 2 shows that while Type II rate has the same shape as the 
instantaneous CSFR, Type Ia rate evolution presents a different shape due to the time delay between the stellar birth and the explosion. This is reproduced at the occurrence of the peak of the SNIa rate which is shifted by a few Gyrs. This characteristic is very important since it provide us with an independent diagnostic of the past history of the star formation rate. Measurement of Type Ia SNe rate at $z \sim 1$ will give us information on the CSFR at much higer redshift. Furthermore, figure 2 shows that for two different models defined by two different 
CSFRs, $R_{Ia}$ has two very distinct shapes (different normalizations and different time  of peak occurrence). Therefore comparison to observed Type Ia rate at$z \sim 1$ would discriminate between models. Whereas Type II SNe are important test for distinguishing between SFR models at only high-z which requires future generation of instruments such as the {\it Next Generation Spatial telescopes}.
In order to compare the model to the observations, we have converted the observed rate in SNu units (SNe/100 yr/ 10$^{10}$ L$_{B\odot}$) into a rate in yr$^{-1}$ using blue luminosities as computed by our code. We find good agreement between our predicted SNe rates and the observations. Although the available measurement of the SNIa rate does not allow one to discriminate between models M1 and M2, the redshift is still low and the statistics is poor. Measurement at $z \sim 1$ with the same statistics, would begin to discriminate between models. This illustrates the fact that the SNI rate at $z = 1$ is sensitive to the CSFR at $z \sim 2 - 3$. 
\section{Constraining the Type Ia progenitors from high-z observations?}
It is not yet clear which of the two main competing models ({\it single degenerate} or {\it double degenerate} model) applies for the SNIa precursors. To investigate how sensitive are our result on the adopted model for $R_{Ia}$, we have used a more general model for Type Ia SNe rate as given by the convenient parametrization introduced by [2] $R_{SNIa}(t)=2.2\times 10^{-15}{\theta}_{SN} t_{15} yr^{-1}$
where $t_{15}$ is the time in units of 15 Gyr. ${\theta}_{SN}$ is a normalisation factor which is 
choosen such as to reproduce the local value. We found that, we still are able to disentangle between the two CSFR at $z=1$. Furthermore we have found that type SNIa at $z=1$ can already be used to put constrains on models for SNIa. We have also found in the frame of this more general model that the best match to observations is obtained for $s>1.5$ models (figure 2) in agreement with [7] who found that $s > 1.4$ in order to account for iron mass to light ratio (IMLR) in clusters of galaxies. 
\section{Conclusion}
We have computed the Cosmic Supernova rates from the CSFR using a self-consistent modelling based on our spectrophotometric evolutionary model. We found that:
-- the adopted standard IMF allows one to fit the observed local and high-$z$ colours of the universe. \\
-- Type II SN rates measurements at $1<z<4$ would allow us to probe the star formation rate while Type Ia rates allow one to probe the past history of the CSFR in the universe.\\
-- our model predictions for SNe rates are in a good agreement with observations locally as well as at higher $z$ and self-consistent with the current limits on the CSFR.\\ 
-- Type Ia measurements at $0<z<1$ would start to put strong constraint on the star formation rate at higher redshift and would discriminate between th models.\\ 
-- current data at $z \sim 0.4$ are not sufficient to disentangle the models of star formation rate, but a better statistics would allow one to discriminate between them and assess whether the $z > 1$ CSFR is higher than directly observed from the UV.\\
-- the comparison between the predicted and the observed Type Ia rates at $z{\sim}1$ would allow to probe the nature of their progenitors. In the frame of more general models, we found that the best match is obtained for models with a steep evolution (i.e $s>1.5$) consistent with the iron content in clusters of galaxies.
  
\acknowledgements{R. Sadat aknowledges support from `` La chaire Blaise Pascal'' }

\end{document}